\documentclass[sigconf,screen,noacm]{acmart}
\AtBeginDocument{%
  \providecommand\BibTeX{{%
    \normalfont B\kern-0.5em{\scshape i\kern-0.25em b}\kern-0.8em\TeX}}}

\setcopyright{none}
\renewcommand\footnotetextcopyrightpermission[1]{}
\usepackage{algpseudocode}
\usepackage[ruled]{algorithm2e}
\newcommand\figref[1]{\figurename~\ref{#1}}

\begin{document}

\title{DID-eFed: Facilitating Federated Learning as a Service with Decentralized Identities
}


\author{Jiahui Geng}
\email{jiahui.geng@uis.no}
\orcid{0000-0002-4205-8230}
\affiliation{%
  \institution{Department of Electrical Engineering and Computer Science,  University of Stavanger}
  \streetaddress{Kjell Arholms gate 41}
  \city{Stavanger}
  \country{Norway}
  \postcode{4021 }
}

\author{Neel Kanwal}
\email{neel.kanwal@uis.no}
\orcid{0000-0002-8115-0558}
\affiliation{%
  \institution{Department of Electrical Engineering and Computer Science, University of Stavanger}
  \city{Stavanger}
  \country{Norway}
}

\author{Martin Gilje Jaatun}
\email{martin.g.jaatun@uis.no}
\orcid{0000-0001-7127-6694}
\affiliation{%
  \institution{Department of Electrical Engineering and Computer Science, University of Stavanger}
  \city{Stavanger}
  \country{Norway}
}

\author{Chunming Rong}
\email{chunming.rong@uis.no}
\orcid{0000-0002-8347-0539}
\affiliation{%
  \institution{Department of Electrical Engineering and Computer Science, University of Stavanger}
  \city{Stavanger}
  \country{Norway}
}

\renewcommand{\shortauthors}{Geng and Kanwal et al.}


\begin{abstract}
We have entered the era of big data, and it is considered to be the "fuel" for the flourishing of artificial intelligence applications. The enactment of the EU General Data Protection Regulation (GDPR) raises concerns about individuals' privacy in big data.  Federated learning (FL) emerges as a functional solution that can help build high-performance models shared among multiple parties while still complying with user privacy and data confidentiality requirements. Although FL has been intensively studied and used in real applications, there is still limited research related to its prospects and applications as a FLaaS (Federated Learning as a Service) to interested 3rd parties. In this paper, we present a FLaaS system: DID-eFed, where FL is facilitated by  decentralized identities (DID) and a smart contract. DID enables a more flexible and credible decentralized access management in our system, while the smart contract offers a frictionless and less error-prone process. We describe particularly the scenario where our DID-eFed enables the FLaaS among hospitals and research institutions.

\end{abstract}





\maketitle
\pagestyle{plain}
\section{Introduction}
With the development of information technology, the world has entered the era of big data. Real-time data on the industrial assembly line is being recorded for production monitoring and management because of the popularization of sensors and smart devices. Individuals are also constantly producing data; their behavior history or consumption on search engines, social networks, shopping sites, and video sites are also recorded. The collected information helps developers analyze user behavior habits and potential needs to improve the user experience or push advertisements. In the past decade, people are enjoying the convenience of data-driven machine learning (ML) techniques in different fields such as autonomous driving~\cite{garcia2019machine}, drug discovery~\cite{lima2016use}, and commodity recommendations~\cite{yin2014commodity}. 
Simultaneously, people are becoming more aware of the importance of data in improving the quality of models; the more prosperous the data, the more accurate and trustable the models are.
However, traditional machine learning methods that require central storage are now facing challenges~\cite{ruj2013decentralized}. On the one hand, scandals like Facebook–Cambridge Analytica Data Scandal have sparked widespread concerns about privacy by the government and the public. GDPR  has been adopted by the EU and becomes enforceable to guide data protection. HIPPA\footnote{\url{https://www.hhs.gov/hipaa/for-professionals/privacy/laws-regulations/index.html}} restricted organizations such as hospitals from sharing users' data for privacy reasons. On the other hand, small-scale companies or research institutions may not get enough scientific research data to build their models and applications. At the same time, the prohibition of data circulation forces the data to exist in isolated data silos maintained by the owners~\cite{xia2017medshare}.

Federated learning (FL)~\cite{wa2020federated} is considered a privacy-preserving machine learning technique to solve the data fragmentation and isolation problem~\cite{hu2015attribute}. FL participants build models collaboratively by sharing the encrypted model parameters instead of the private data. 
Although FL has been widely studied and used in reality,  there is very little related work documenting that FL is provided to interested third parties as a service. There are already many cloud-based machine learning as a service (MLaaS)~\cite{philipp2020machine} services, and they are primarily profitable through the provision of computing resources or APIs. However, they don't support collaborative learning, and users need to upload their data and hence lose control of the data. Federated learning as a service (FLaaS)~\cite{kourtellis2020flaas} has a more general, broader usage scenario where the data user is not the data owner~\cite{gruner2019comparative}. With FLaaS, there is no data transfer, and the owner can keep the data private and secure. In contrast, developers and data scientists can focus on algorithm development and do not need to bother with data collection.

In this paper, we present a Federated Learning Service system (DID-eFed) that encompasses the construction of a Federated Learning System and the provisioning and access management of services. We design a service access control based on decentralized identity (DID), a claim-based distributed identity management system built on blockchain technology. The verifiable claim mechanism will help make trusted authorized access. DID is a user-centric identity model where each user intelligently applies for a digital identity but has multiple claims, which will prevent misuse of the service and Sybil attacks~\cite{fung2018mitigating}. We also take advantage of smart contracts to make the service access management process smoother. The execution of automated smart contracts will monitor the entire process, reducing and ultimately eliminating errors associated with manual review and response to prior authorization requests and reducing appeals due to misinterpretation of manually written authorization forms.

The major contributions of this paper are described as follows:
\begin{enumerate}
    \item We propose a FLaaS system DID-eFed in which the FLaaS is facilitated with the help of Smart Contract and DID. We explain how we enable the FLaaS to users and how we perform effective permission and privacy management. 
    \item We describe a scenario, where DID-eFed will be a feasible solution to enable the FL among the hospitals and research institutions. 
    \item We discuss the benefits of DID-eFed and point out some future research directions 
\end{enumerate}


The remainder of this paper is structured as follows: We provide relevant background in section~\ref{sec:prelim}, and details on related work are discussed in section~\ref{sec:related}. 
Section~\ref{sec:approach} 
describes our proposed approach followed by challenges that are discussed in ~\ref{sec:challenges}. Finally, Section~\ref{sec:conclusion}  concludes and describes future prospects related to the proposed system. 

\section{Preliminaries}
\label{sec:prelim}
In this section, we will present preliminaries and related background knowledge on FL, Blockchain,  Smart Contract, and DID.
\subsection{Federated Learning}
FL is an emerging ML paradigm first proposed by Google~\cite{konevcny2015federated} in 2016, where participants collaboratively train a global model by exchanging the encrypted model parameters instead of private data. 

\subsubsection{Categories of Federated Learning}
Yang et al.~\cite{yang2019federated} classified FL into horizontal federated learning (HFL), vertical federated learning (VFL), and federated transfer learning (FTL) according to the data distribution. HFL refers to the case where datasets owned by different participants share similar feature space, but the sample ID differs. VFL deals with the scenarios in which participants have significant overlaps in the sample ID but differ in the feature space. Under FTL, participants have little overlap in both the sample ID and the feature space. It is worth noting that, if not emphasized, FL refers to HFL by default.

\subsubsection{Training Process of Federated Learning}
The client-server architecture is a typical framework for FL~\cite{zhang2019efficient}. In this architecture, $N$ clients with the same features collaboratively train a machine learning model with the help of a server. Let $D$ denote the total training data and $D_i$ denote the local data held on client $C_i$, where ${i \in \{1..N\}}$. The objective of FL is to  minimize the global loss $l(\omega, D)$ over all training samples~\cite{zhao2020local}.
\[\omega^* = \arg\min_{\omega} {l(\omega, D)} = \arg\min_{\omega} \sum_{i=1}^N l(\omega, D_i)\]
Taking into consideration the communication overhead and the stability of the network, the FedAvg algorithm~\cite{li2019convergence} is proposed and FL is simplified to that the server aggregates models from a subset $S_t$ of $K$ clients at the $t$-th iteration.  
\[\omega^{t+1} = \sum_{i} p_{i, t} \cdot \omega_{i, t}\]
where $i \in S_t$, $p_{i, t}$ is the weight of corresponding model and $\sum_i p_{i, t} = 1$.

\begin{algorithm}[t]
	\caption{Federated Averaging (FedAvg)}
	\label{alg:algorithm1}
	\KwIn{${T}$:total iterations, ${N}$: total clients, $K$: subset size, $E$:local training epochs, $\omega^0$: initial model}
	\KwOut{Final global model $\omega^{T+1}$}  
	\BlankLine
	Send the initial global model $\omega^0$ to all clients\\
	\For{\textnormal{t = 1,...,T }}{
	    server select a subset $S_t$ of $K$ devices randomly\\
	    \For{i $\in$ $S_t$}{
	        client i trains model on local data for $E$ epochs \\
	        client i send local model $\omega_{i, t}$ to the server \\
	    }
	    server aggregates $\omega^{t+1}$ as $\omega^{t+1} = \sum_{i} p_{i, t} \cdot \omega_{i, t}$
	}
\end{algorithm}

As summarized in Algorithm 1, at each iteration $t$, a subset of clients will be selected to  train the local model for $E$ epochs after which the local models will be aggregated on the server.

\subsection{Blockchain}
The first practical implementation of blockchain was made by Satoshi Nakamoto in 2008 to serve as the public transaction ledger of the cryptocurrency Bitcoin~\cite{nakamoto2008peer}.  A blockchain is essentially a growing list of linked blocks. Blocks are added to the blockchain by consensus among the majority of nodes in the system. Each block contains a timestamp, transaction data (typically represented in a Merkle tree), and the cryptographic hash of the previous block. In this way, the blocks are linked together in chronological order. The cryptographic hashing algorithm ensures that the transaction data in each block is immutable and the linked blocks in the blockchain cannot be tampered with~\cite{wang2018blockchain}.


\subsection{Smart Contracts}
A smart contract is an automatically executed agreement subject to its explicit terms and conditions. The agreement stores and enforces the terms of the contract on the blockchain~\cite{yasin2016online}.
Smart contracts allow for the execution of reliable transactions without the intervention of a third party. With the implementation of Ethereum~\cite{Wood2014ETHEREUMAS} in 2015, blockchain-based smart contracts are for general-purpose computing executed on a blockchain or distributed ledger. When a smart contract meets predetermined terms and conditions, it is automatically executed according to the rules. A simple example might be life insurance. The terms of the policy would be encoded into the smart contract. If the policy holder passes away, a notarized death certificate would be provided as the input trigger for the smart contract to release payment to the designated beneficiary~\cite{luu2016making}.

\subsection{Decentralized Identity (DID)}
DID is essentially a claim-based identity system supported by blockchain technology~\cite{bouras2020distributed}.  A blockchain-based distributed ledger replaces the centralized authority as the trusted source. The identity information itself is not stored in the ledger but in a wallet managed by the user. By controlling what information is shared from the wallet to the requesting third party, users are able to manage their identity and privacy better online~\cite{avellaneda2019decentralized}.

\subsubsection{Claim-based Identity} 
Claim-based identity is a method of authenticating a user, application or device with another system that abstracts specific information about the entity,  providing relevant authorization for access management.

Claims-based identity removes the need for applications to perform authentication tasks, making identity management with less  effort possible.  Multiple claims of identity can help avoid frequent authorization requests for each access of an application. It also enables the federation of identities that the external users can access using their own identities without creating new accounts. Claims-based identity offers more versatility when the system requires unique attributes as claims for access~\cite{luecking2020decentralized}.

\subsubsection{DID Components}
    \begin{itemize}
    \item {\bf Decentralized Identifier (DID Identifier)}
    A DID identifier is a new type of identifier that enables verifiable, decentralized digital identity. It is a simple text string consisting of three parts: 1) the DID URI scheme identifier, 2) the identifier of the DID method, and 3) the identifier specific to the DID method.

    \item {\bf DID Document}
    The DID Document is usually a JSON document that contains public key material, authentication
descriptors, and service endpoints. It contains verifiable claims describing the identity.  The DID Document enables a DID controller to prove control of the DID. In short, the DID Identifier is the Identifier of the corresponding DID Document, and the DID Document contains information such as what the DID can authorize and in which services the DID can be used.

    \item {\bf DID Resolver and Driver}
    A DID Resolver is a server that uses a collection of DID drivers to provide a standard means of querying and resolving DID Identifiers across decentralized systems, and returns the DID document associated with the DID Identifier. When a DID Identifier is passed to the DID Resolver, the resolver uses the appropriate driver to interface with the decentralized system and retrieve the matching DID document.
    \end{itemize}

\section{Related Work}
\label{sec:related}
Along with the development of machine learning and cloud computing technologies, a novel service: machine learning as a service (MLaaS), has emerged~\cite{philipp2020machine}. MLaaS is a set of services provided by cloud service providers that offer ready-made, slightly generic machine learning tools that any organization can adopt as a part of their working needs. All the training data need to be uploaded to the MLaaS providers like AWS Machine Learning \footnote{\url{https://aws.amazon.com/machine-learning}}, Google Cloud AI 
\footnote{\url{https://cloud.google.com/ai-platform}} and Azure Machine Learning \footnote{\url{https://azure.microsoft.com/en-us/services/machine-learning/}}. The MLaaS providers will manage it and perform machine learning tasks that require high hardware performance. However, the existing MLaaS does not support federated learning, and the centralized storage of cloud services is contradictory to the distributed storage in federated learning scenarios~\cite{sun2019can}.

The introduction of the concept of federated learning has attracted extensive research and attention from academia and industry. Many well-known communities have proposed open-source federated learning frameworks to help data scientists quickly verify federated learning algorithms and support large-scale deployment of federated learning tasks for commercial purposes, for example, FATE\footnote{\url{https://github.com/FederatedAI/FATE}}, PySyft\footnote{\url{https://github.com/OpenMined/PySyft}}, PaddleFL\footnote{\url{https://github.com/PaddlePaddle/PaddleFL}}, FedML\footnote{\url{https://github.com/FedML-AI/FedML}}. These frameworks can be used as the basis for building our federal services. 

There are two articles with similar ideas to ours, but the specific implementation methods are different. Jose et al.~\cite{jose2019totem} proposed a token-based computing system that combines both blockchain technologies and big data systems (Hadoop), smart contract pre-checks of the user task code, and monitoring the resource availability and the running time. Kourtellis et al.~\cite{kourtellis2020flaas} introduced their architecture of a FLaaS, but they only designed the interface APIs without considering the system security and access control. 

After the advent of blockchain~\cite{nakamoto2008peer}, it has been considered as a better solution for security for its decentralization, immutability, transparency, and security properties~\cite{mendis2020blockchain}. Besides, smart contact based access control helps turn the workflow into an automated process because of self-executing, self-verifying, and tamper-resistant characteristics. Smart contract based access control for data sharing is attracting more attention from researchers. In MeDShare~\cite{xia2017medshare}, data transfer from one entity to another, and all operations performed within the system are recorded in a tamper-proof manner based on the blockchain technique. Wang et al.~\cite{wang2018blockchain} proposed a blockchain-based framework for data sharing with fine-grained access control in decentralized storage systems. However, the vast majority of solutions that use smart contracts are role-based access controls. Some researchers are working on using decentralized identity, an open, trustworthy, interoperable, and standards-based identity management ecosystem to boost the trust between the organization and its customers and partners~\cite{reed2016requirements}.

Many studies combine federated learning with blockchain or DID, but their work often relies on blockchain to implement the incentive system, sharing system, or building reputation system for federated learning ~\cite{fan2020diam} ~\cite{volkov2020addressing}.
Mendis et al.~\cite{mendis2020blockchain} proposed a decentralized, secure FL framework, which uses blockchain to record the contribution of data owners. Luecking et al.~\cite{luecking2020decentralized} used a similar DID technology to ours, but their focus is on building a trust mechanism for the IoT system.
Harris et al.~\cite{harris2019decentralized} described a framework for sharing and improving a machine learning model where anyone can freely access the model's predictions or provide data to help improve the model.

\section{Proposed Approach}
\label{sec:approach}
In this paper, we propose DID-eFed, which facilitates federated learning with DID to enhance the system security, user privacy and ease of operation. 
\subsection{Scenarios}
The cooperation between hospitals and research institutions will be a typical scenario for our proposed approach. Data sharing is limited, and medical analysis is expected to be performed locally instead of off-site.
Health data is susceptible to misuse, and its usage is strictly regulated. Even with the help of anonymization techniques, there remains the risk of privacy leakage. Also, it takes considerable time, money, and human effort to collect, organize, and maintain high-quality datasets. Consequently, such datasets contain a significant business value, and the data owners are not willing to share them freely. 
 
Successful implementation of FL may have great potential for large-scale applications of precision medicine, leading to unbiased decision-making, sensitivity to rare diseases that best reflects the individual's physiological condition while respecting governance and privacy concerns~\cite{lyu2020threats}. In this scenario, the hospitals will actively provide health data and computational resources for their partners to conduct federated learning for medical analysis. Considering the security of the overall system, only verified users are allowed to perform computing tasks within such a federated service. 
 
\subsection{Architecture}
\subsubsection{Federated Learning System}
The client is a node with training data, and the server is a node that maintains good communication with all clients. The server controls the startup and execution of the FLaaS, and is connected to the access management system. The framework of federated learning will be packaged as a mirror image and installed on all nodes. When the FLaaS is configured and launched, all the clients train the models with the same structure, and after multiple rounds of training on the client data, updated local models will be sent to the server to update the global model. Both our server and client have a high-availability topology configuration. Once a node, whether it is a server or a client, fails, the standby node will be started and added to the training of federated learning. The Federated Learning System provides two API, as follows:\par
\begin{itemize}
    \item Data API
    \begin{itemize}
        \item allows users to view the metadata of all clients, including data type, data features and labels, data size, and data distribution.
        \item allows users to view part data samples, including data visualization, image display.
        \item allows users to customize the data filters, data features, and data sampling strategies used for training.
    \end{itemize}
    \item Model API
    \begin{itemize}
        \item allows user to customize the training model, including initialization, optimizer and loss function, etc.
        \item allows user to customize the hyper-parameters for FL, including the numbers of total iterations, local epochs, total clients, subset clients, model compression method and model aggregation strategy.
    \end{itemize}

\end{itemize}

\subsubsection{DID-based Access Management System}
\begin{itemize}
    \item The entire access system contains claim holder, claim issuer and claim verifier.
    \begin{itemize}
        \item Claim holder is a user who wants to access the FLaaS. He will first need to register a DID, which is unique to the individual in our system. DID applications may require the fundamental authentication information. This ensures the authenticity of the identity and prevents Sybil attacks~\cite{maram2020candid,xie2019dba}. There exists several open-source DID frameworks such as  Hyperledger Indy~\cite{dhillon2017hyperledger} and uPort~\cite{jacobovitz2016blockchain}; these blockchain-based solutions support the mutual recognition of identities in different DID systems and help to dovetail our FLaaS into the larger ecosystem. Then the user needs to apply to the claim issuer to obtain a claim. If the user can prove that he is a member of the consortium or a paid external user through his profile, he will become a claim holder.
        \item Claim issuer creates a verifiable claim associating with a specific subject, and transmits it to the claim holder. Example issuers include governments, organizations, associations, and corporations.
        In principle, the claim issuer should be the first person to appear in the entire system to issue claims to other users, but this is often not the case. Therefore, the access system needs to take into consideration legacy compatibility. When certain services have been launched, and accounts have been created to grant access, our solution should be compatible with these accounts and translate these accounts into a claim.
        \item Claim verifier performs by requesting and receiving a verifiable claim that proves the claim holders possess the required verifiable claims according to the access rules. The data owner is accountable for who has access to information assets and customizes the regulations by the Access APIs. Here we turn claim verification into an automated process with the help of smart contract technology.

    \end{itemize}

\end{itemize}
\begin{itemize}
    \item Access API
    \begin{itemize}
        \item allows users to view access permissions, i.e., claims that need to be provided.
        \item allows users to view how to use FLaaS and access addresses for example data visualization and training process visualization for analysis.
    \end{itemize}
\end{itemize}

\subsection{Workflow}
\subsubsection{System Setup}
FLaaS Provider deploys clusters to form a highly available network composed of clients and servers, and the server maintains communication with all clients. Data will be stored in a specified location for the user to access according to the provided API. The current FLaaS consists of a set of Docker containers configured and managed by Docker Compose.
To deploy the DID identity system; first, the consortium and the users who wish to access the FLaaS register a DID according to their profile. The consortium, made up of representatives of the data's contributors and providers of the FLaaS, is eligible to become claim issuers. When the user meets the conditions, the DID Identifier will be sent to the user. The corresponding claim will be saved in the corresponding DID document and sent to the distributed ledger as a trusted part.

\begin{figure}[h]
  \centering
  \includegraphics[width=0.9\linewidth]{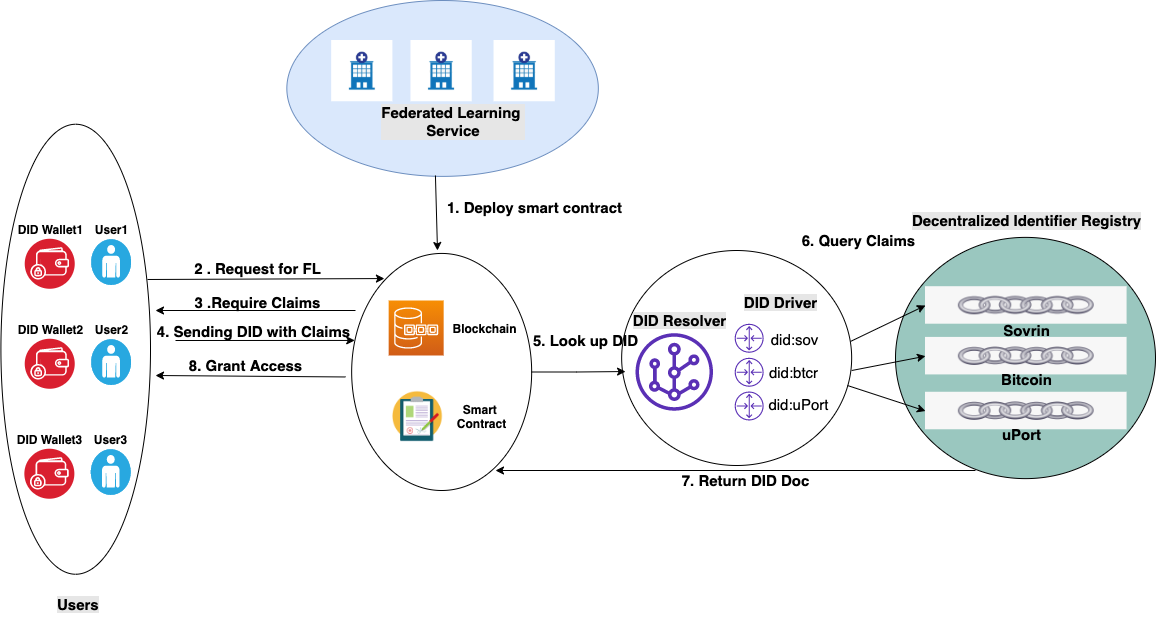}
  \caption{Workflow: smart contract requests to lookup the DID  }
  \label{fig2}
\end{figure}

\begin{figure}[h]
  \centering
  \includegraphics[width=0.9\linewidth]{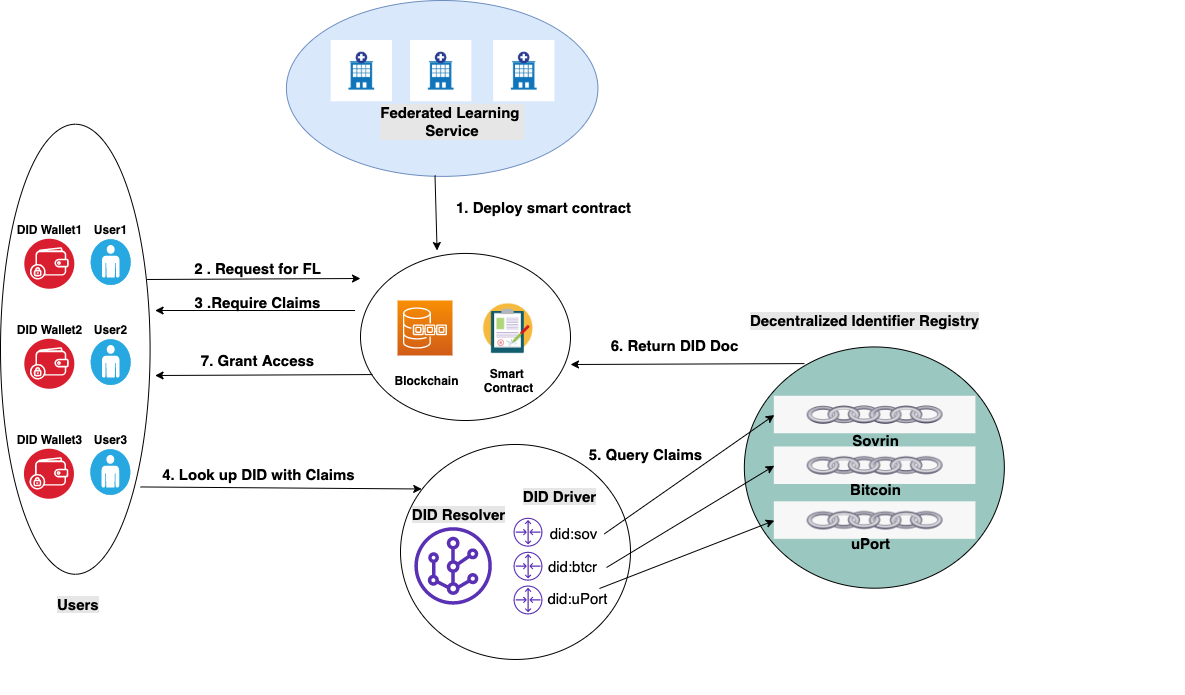}
  \caption{Workflow: user requests to lookup the DID}
  \label{fig3}
\end{figure}

\subsubsection{Access Management} 
The workflow involves the deployment of the smart contract, the authentication of the DID, and the grant of the FLaaS. 
For the authentication part, we design and discuss two different schemes, as illustrated in \figref{fig2} and  \figref{fig3}. We will first describe the details in the first scheme, then we explain the differences between the schemes; finally, we analyze the pros and cons respectively.

We assume that all necessary information about claims is stored in the digital wallet and the decentralized identifier registries, and thus we skip the discussion about DID registration. A group of data owners will provide the FLaaS, they define the resource access policy, and the required claims. Any qualified user can get access to the FLaaS after the authentication of the DID system.

The corresponding descriptions of step numbers in~\figref{fig2} are shown as follows.

\textcircled{1} Data owners deploy smart contracts about access policy to the blockchain network. Different owners can have different access policies which depend on their own data management regulations. Data owners can also provide different FLaaS for users with various research interests.  
\textcircled{2} An interested user needs to query the blockchain for the authentication information of the FLaaS.
\textcircled{3} Once the request is successful, the user will get the required service-specific claims. 
\textcircled{4} The user will lookup the DID with claims. \textcircled{5} After verifying the user DID, the DID resolver will look up the DID to its DID Document with the help of DID drivers\textcircled{6}, which knows how to connect to the target system. \textcircled{7} The content of DID Documents will be parsed and the smart contract will be executed to  verify the required claims.  \textcircled{8}  If predetermined conditions have been met and verified, meta-data about the training data access will be granted to the user.  The meta-data about the training data and computational requirements enable the users to better understand the data properties and change their algorithm correspondingly. The blockchain is then updated when the transaction is completed.

As illustrated in ~\figref{fig3}, the user will lookup the DID directly instead of the smart contract in the second scheme. We argue that in both schemes, the DID document will not be sent to the user directly since the users can modify the DID Document on purpose. 
For the first scheme, the DID Resolver stays behind the blockchain system, the user only needs to communicate with the blockchain. However, the blockchain platform may suffer from DoS or DDoS attacks. After requiring the claims to the user, the smart contract keeps listening to the response, while the user can keep sending fake DID information to interfere with the normal operation of blockchain services. The memory of the blockchain platform is limited since all transactions need to be verified, packaged, and sorted in the memory pool before being sent to the blockchain network by miners. 

For the second scheme, the user needs to establish additional communication with the DID resolver, and the risk of attack is transferred to DID Resolver. However, the DID Resolver should have enough memory for recommended practices for preventing attacks.

\subsubsection{FLaaS} 
When a FL task is submitted, it enters the task queue. The task scheduling engine optimizes the order of execution based on the resources required, estimated by the user to make the system as efficient as possible. When the task starts, the user will receive a notification about the visualization address to be aware of the changes in the various metrics during the training process, and a link to the download of the model after the training is complete.
Corresponding images will be automatically pulled on the servers and clients. The data volumes on the target machine will be mounted to the containers according to the meta-data information. At the same time, the user's training script will be placed in the container to be started and executed. When the docker containers on all nodes are ready, the FL algorithm will proceed. The FL will initialize the same models on different clients. After training for the same epochs, the updates of the models will be sent to the servers, which will be aggregated with various strategies.

\section{Challenges}
\label{sec:challenges}
There are remaining challenges regarding workflow and load balancing, privacy, analysis tools, and cloud platforms. 

\subsection{Workflow Schedule \& Load Balance}
In a federated learning system there are many users working on federated learning at the same time, and if task scheduling is not optimised, the whole system becomes inefficient in terms of resource utilisation. This requires a combination of algorithms to schedule federated learning tasks based on user resource usage, CPU and GPU usage, and model runtime to reduce the time spent on all tasks. Load balancing can be performed to even out the pressure on the servers and clients and reduce communication blocking~\cite{geyer2017differentially}. In addition, topology improvements and model compression can be made to improve system efficiency. In addition to this, the learning task is terminated early if the federal learning is monitored for non-convergence, or if training does not improve on key metrics.

\subsection{Privacy-preservation \& Attack Defense}
Although users do not need to upload data to a central server in FL training and there is no communication between different clients, there is still a risk of compromising user privacy. Many researchers consider using differential privacy (DP) ~\cite{seif2020wireless,truex2020ldp,wei2020federated}, homomorphic encryption (HE)~\cite{chai2020secure,hardy2017private}, or multi-party computation (MPC)~\cite{yao1982protocols} in the FL system to enhance data security. These methods would either compromise the accuracy of the training model or increase the computational load and make the computation process slower. Additionally, FL needs to take into account malicious attacks~\cite{li2020learning} or even unintentional ones. As FL cannot look at all users' data, if the users are of low quality and make serious labeling errors, this can lead to poor performance of FL. Additionally, backdoor attacks~\cite{bagdasaryan2020backdoor,xie2019dba} can be a concern. We can continue our work using ML techniques and blockchain technology to detect abnormal participants and build reputation or incentives.
\subsection{Analysis Tools for FL}
DID-eFed is a distributed system, and we need analysis tools to help us identify problematic nodes, such as a visualization platform like Tensorboard~\cite{rampasek2016tensorflow} or other performance profiling tools.

\subsection{Cloud Platform}
Many users currently store data on different cloud platforms like AWS, GCP and Azure, and we need to develop frameworks to support federated learning between various cloud platforms in the future.

\section{Conclusion}
\label{sec:conclusion}
Federated learning provides a solution to improve user privacy because most personal data is stored on private devices. In this paper, we have presented an architecture to establish trustworthy federated learning. The novel architecture consists of three components; a DID system, smart contract, and a FLaaS.  DID is utilized to enable complete control over personal identity independent of any certification authority.  Users who are interested in the FLaaS can get access with the required claims without leaking irrelevant private information. 
The authentication is executed by smart contract in an immutable manner.  
Within this architecture, data owners are able to keep the data secure and at the same time provide a FLaaS to users to perform specific learning tasks.  The smart contract plays the role of system monitor based on a blockchain platform, it prevents many malicious attacks and enhances the management of FL automatically.

We have discussed the performance and the scalability of the system, analyze the risks and attacks that can exist in the system, and possibility of considering a reward for high-quality data providers or algorithm providers in conjunction with the blockchain.

\begin{acks}
This research is supported by CLARIFY Project. CLARIFY is European Union’s Horizon 2020 research and innovation program under the Marie Skłodowska-Curie grant agreement No. 860627.\par
\textbf{Disclaimer clause for CLARIFY (H2020-MSCA-ITN-2019)}.\par
The results of this publication reflect only the authors' view and the Commission is not responsible for any use that may be made of the information it contains.
\end{acks}
\bibliographystyle{ACM-Reference-Format}
\bibliography{DID.bib}
\end{document}